\documentclass[english,aps, prl, superscriptaddress,twocolumn,showpacs]{revtex4}
\usepackage{mathptmx}
\usepackage[T1]{fontenc}
\usepackage[latin9]{inputenc}
\usepackage{graphicx}

\makeatletter

\usepackage{babel}
\makeatother

\begin{document}

\title{Unraveling the {}``Pressure-Effect'' in Nucleation}

\author{Jan Wedekind}

\email{janw@ffn.ub.es}

\affiliation{Departament de Física Fonamental, Universitat de Barcelona, Martí
i Franquès 1, 08028 Barcelona, Spain}

\author{Antti-Pekka Hyvärinen}

\affiliation{Finnish Meteorological Institute, Erik Palménin aukio 1, P.O. Box
503, F1-00101 Helsinki, Finland}

\author{David Brus}

\affiliation{Finnish Meteorological Institute, Erik Palménin aukio 1, P.O. Box
503, F1-00101 Helsinki, Finland}

\affiliation{Laboratory of Aerosol Chemistry and Physics, Institute of Chemical
Process Fundamentals, Academy of Sciences of the Czech Republic, Rozvojová
135, CZ-16502 Prague 6, Czech Republic}

\author{David Reguera}

\affiliation{Departament de Física Fonamental, Universitat de Barcelona, Martí
i Franquès 1, 08028 Barcelona, Spain}

\begin{abstract}
The influence of the pressure of a chemically inert carrier-gas on
the nucleation rate is one of the biggest puzzles in the research
of gas-liquid nucleation. Different experiments can show a positive
effect, a negative effect, or no effect at all. The same experiment
may show both trends for the same substance depending on temperature,
or for different substances at the same temperature. We show how this
ambiguous effect naturally arises from the competition of two contributions:
nonisothermal effects and pressure-volume work. Our model clarifies
seemingly contradictory experimental results and quantifies the variation
of the nucleation ability of a substance in the presence of an ambient
gas. Our findings are corroborated by results from molecular dynamics
simulation and might have important implications since nucleation
in experiments, technical applications, and nature practically always
occurs in the presence of an ambient gas.
\end{abstract}



\maketitle
A gas phase can be supersaturated considerably beyond its equilibrium
condensation point before liquid drops form spontaneously. The pathway
of the phase transition is blocked because microscopic droplets are
thermodynamically less favorable than the bulk vapor. Therefore, the
transition can only be initiated by rare fluctuations exceeding a
critical size, called the \emph{critical nucleus}. The formation of
such a critical nucleus is the limiting step in the transition and
its frequency of occurence is called the \emph{nucleation rate}. Nucleation
is behind most phase transition and plays a crucial role in atmospheric
processes such as the formation of aerosols or the condensation of
water vapor into clouds \cite{Kulmala,turnved}. An accurate experimental
evaluation of the nucleation of atmospherically relevant substances
and its correct theoretical prediction are essential for a better
understanding of climate change and are the subject of intense investigations
\cite{Winkler}.

Nucleation is highly sensitive to small changes in the state variables
describing the system, most notably to temperature. However, condensation
is inevitably connected with the release of latent heat. In experiments
this latent heat is removed by the presence of a large background
of carrier-gas. This carrier-gas should be noncondensing and chemically
inert and should have no influence on the nucleation except serving
as the desired heat bath. Nevertheless, many experimental results
suggest that there can be an influence on the nucleation rate that
can span some orders of magnitude, depending on the pressure and type
of the carrier-gas (see Refs.\,\cite{Brus_et_al,Brus_Accep} and
references therein). Unfortunately, the overall picture is far from
clear. Comparisons of experimental findings cover all possibilities:
no effect, increase, or decrease of the nucleation rate with carrier-gas
pressure. There were some concerns if the pressure-effect is not just
an experimental artifact, which cannot be completely ruled out in
all cases. However, continuous improvements in the experimental setups
rather confirmed than remedied this undesireable and elusive effect.

Following the experimental evidence, theoretical works have tried
to explain the pressure-effect with varying success. These approaches
include changes to classical nucleation theory (CNT) accounting for
nonideal behavior of the vapor and/or carrier-gas \cite{Ford}, treating
the carrier-gas as not fully inert or the problem as binary nucleation
\cite{Oxt_Laak,Luij_vDong,Kalikm} or analyzing its influence on
cluster stability and the impingment rate \cite{Nov_Reiss}, and
many more. There remain many contradictions and differences both in
the direction and in the magnitude of the effect. These are complicated
by the discrepancies between experimental results themselves and in
the comparison of experiment and theory. 

Here we present a simple yet physically very appealing model that
resolves many of the apparent contradictions of the pressure-effect.
We take a deliberate step back and incorporate the presence of a carrier-gas
into classical theory in a most natural manner that accounts for the
two primal contributions of the carrier-gas: the efficiency of thermalization
and the additional work that a cluster has to spend for growing in
its presence. These contributions have opposite trends and we show
how this may be responsible for the existence of apparently contradictory
results.

Nucleation theory \cite{Debene} usually aims at calculating isothermal
nucleation rates, taking the constant value of the temperature and
the idea of a non-influential carrier-gas for granted. In CNT, the
work of formation of a droplet of size $n$ at constant pressure $p$
and temperature $T$ is \cite{Debene}\begin{equation}
\Delta G\left(n\right)=-n\Delta\mu+s_{1}\gamma n^{2/3},\label{eq:work-of-formation}\end{equation}
i.e., the combination a volume term related to the difference in chemical
potential $\Delta\mu$ between the vapor and the liquid and a surface
term needed to build the liquid interface $A=s_{1}n^{2/3}$ with the
surface tension $\gamma$ ($s_{1}=\left(36\pi v_{l}^{2}\right)^{1/3}$
is the surface area per monomer and $v_{l}$ is the volume per molecule
in the bulk liquid). The vapor pressure $p$ is assumed as ideal and
the liquid cluster is considered as an incompressible spherical drop
with a sharp interface and bulk liquid properties. The free energy
has a maximum at the critical size $n^{*}$ and its height $\Delta G^{*}$
is\begin{equation}
\Delta G^{*}=\frac{16\pi}{3}\frac{v_{l}^{2}\gamma^{3}}{\Delta\mu^{2}}.\label{eq:CNT_barrier}\end{equation}
The isothermal steady-state nucleation rate then is\begin{equation}
J\mathrm{_{CNT}}=K\exp\left(\Delta G^{*}/k\mathrm{_{B}}T\right),\label{eq:CNT_rate}\end{equation}
where $K$ is a kinetic prefactor. In CNT, the difference in the chemical
potential is given by\begin{equation}
\Delta\mu=k\mathrm{_{B}}T\ln S-v_{l}\left(p-p\mathrm{_{eq}}\right).\label{eq:chem_pot}\end{equation}

Here, $S=p/p\mathrm{_{eq}}$ is the supersaturation and $p\mathrm{_{eq}}$
is the equilibrium vapor pressure. The second term in Eq.\,(\ref{eq:chem_pot})
is arising from the pressure-volume work the liquid drop has to perform
against the ambient \emph{vapor }pressure \cite{Reg_small}. It is
typically small and hence commonly neglected but we keep it here for
clarity. We now naturally account for the presence of an ideal carrier-gas
by noting that the cluster must also perform $pV$-work against the
ambient \emph{carrier-gas} pressure, $W\mathrm{_{c}}=p\mathrm{_{c}}V(n)=nv_{l}p\mathrm{_{c}}$
\cite{Bowles_2}. The work of formation including $pV$-work now
reads\begin{equation}
\Delta G_{pV}\left(n\right)=-n\Delta\mu\mathrm{_{eff}}+s_{1}\gamma n^{2/3},\label{eq:pv_work}\end{equation}
where we have cast the $pV$-contributions into an {}``effective
chemical potential''\begin{equation}
\Delta\mu\mathrm{_{eff}}=k\mathrm{_{B}}T\ln S-v_{l}\left(p+p\mathrm{_{c}}-p\mathrm{_{eq}}\right).\label{eq:eff_chem_pot}\end{equation}
Replacing $\Delta\mu$ for $\Delta\mu\mathrm{_{eff}}$ in Eq.\,(\ref{eq:CNT_barrier}),
the barrier becomes $\Delta G_{pV}^{*}=\left(16\pi/3\right)\left(v_{l}^{2}\gamma^{3}/\Delta\mu\mathrm{_{eff}}^{2}\right)$
and the $pV$-corrected rate follows directly from Eq\@.\,(\ref{eq:CNT_rate})
as\begin{equation}
J_{pV}=K\exp\left(\Delta G_{pV}^{*}/k\mathrm{_{B}}T\right).\label{eq:pV_rate}\end{equation}

The main designated role of a carrier-gas is to keep $T$ constant.
However, this thermalization might not be perfect and nucleation then
happens under nonisothermal conditions. The classical work of Feder
\emph{et al}. offers analytical expressions to quantify this influence
on the nucleation rate \cite{Feder,thermostats}. Physically, it
is controlled by the competition between the energy increase due to
latent heat and the energy removal through elastic collisions with
vapor and carrier-gas molecules. The parameter\begin{equation}
q=h-\frac{k\mathrm{_{B}}T}{2}-\gamma\frac{\partial A(n)}{\partial n}\label{eq:q_parameter}\end{equation}
quantifies the energy released when a monomer is added to a cluster,
which is the latent heat $h$ per molecule (corrected by a small amount
$k\mathrm{_{B}}T/2$) minus the energy spent on increasing the surface
area $A(n)$ against the surface tension. The mean squared energy
fluctuation removed by collisions with impinging molecules is\begin{equation}
b^{2}=2k\mathrm{_{B}}^{2}T^{2}\left(1+\frac{N\mathrm{_{c}}}{N}\sqrt{\frac{m}{m\mathrm{_{c}}}}\right)\label{eq:b_squared}\end{equation}
for an ideal monatomic vapor and carrier-gas \cite{thermostats}
($m$ is the molecular mass, $N$ the number of molecules of the condensable).
Eq.\,(\ref{eq:b_squared}) indicates that a large number of carrier-gas
molecules $N\mathrm{_{c}}$ and a light (small molecular mass $m_{\mathrm{c}}$)
carrier-gas are most effective for a good and fast thermalization.
Finally, the influence of nonisothermal effects on the steady-state
nucleation rate is determined by the combination of $q$ and $b$:
\begin{equation}
J_{\mathrm{nonisoth.}}=\frac{b^{2}}{b^{2}+q^{2}}J\mathrm{_{isoth.}}.\label{eq:nonisotherm_rate}\end{equation}

\begin{figure}
\includegraphics[width=0.95\columnwidth]{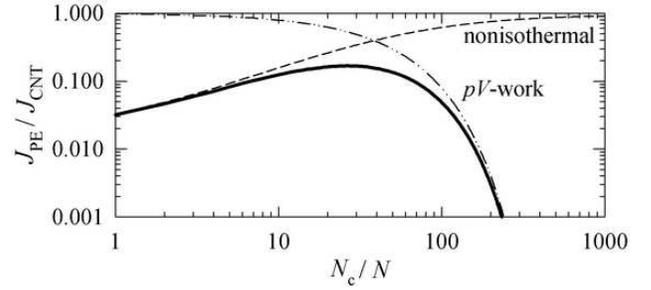}

\caption{\label{fig:concept}Deviation of the nucleation rate from CNT due
to the full pressure-effect, Eq.~(\ref{eq:PE_deviation}), as a function
of the ratio of carrier-gas over vapor molecules (solid) arising from
the two contributions of nonisothermal effects (dash, Eq.~(\ref{eq:nonisotherm_rate}))
and $pV$-work (dash-dot, Eq.~(\ref{eq:pV_rate})) for Argon at 50\,K
and $S=869$.}

\end{figure}
We now obtain the full pressure-effect (PE) of the carrier-gas on
the nucleation rate by combining Eqs.\,(\ref{eq:nonisotherm_rate})
and (\ref{eq:pV_rate}), taking the latter as the isothermal but $pV$-corrected
rate: \begin{equation}
J_{\mathrm{PE}}=\frac{b^{2}}{b^{2}+q^{2}}J_{pV}.\label{eq:PE_rate}\end{equation}

Obviously, the magnitude of the rate calculated by Eq.\,(\ref{eq:PE_rate})
depends on the estimate of the underlying CNT, which often can be
off by many orders of magnitude \cite{Iland,merikPRL}. In our case
however, we are only interested in the \emph{deviations} arising from
the pressure-effect. By renormalizing Eq.\,(\ref{eq:PE_rate}) we
get a reasonable and mostly model-independent estimate of this deviation
\cite{McGraw_Lett}:\begin{equation}
\frac{J_{\mathrm{PE}}}{J_{\mathrm{CNT}}}=\frac{b^{2}}{b^{2}+q^{2}}\frac{J_{pV}}{J_{\mathrm{CNT}}}.\label{eq:PE_deviation}\end{equation}

Fig.\,\ref{fig:concept} shows the change of the rate as a function
of the ratio of carrier-gas to vapor molecules, $N_{\mathrm{c}}/N$,
which for perfect gases is the same as the ratio of carrier-gas over
vapor pressure, $p_{\mathrm{c}}/p$. The calculation was performed
for argon at 50\,K and a supersaturation of $S=869$ \cite{Wede_rates}
and the latent heat $h$ was calculated via the Clausius-Clapeyron
relation. Fig.\,\ref{fig:concept} also shows the individual contributions
coming from nonisothermal corrections, which always have a positive
effect on the rate, and $pV$-work, whose effect is always negative.
The competition of both terms first leads to an increase in the rate
with increasing carrier-gas pressure due to better thermalization.
Then, the penalty of \emph{pV}-work takes over and leads to an overall
decrease of the rate. 

We have performed standard molecular dynamics (MD) simulations of
Lennard-Jones (LJ) argon nucleation at 50\,K ($N_{\mathrm{Ar}}=343$,
$S=869$) and 80.7\,K ($N_{\mathrm{Ar}}=512$, $S=10.5$) with LJ-helium
as carrier-gas in order to verify the predictions of our model. Details
of the simulation are similar to Ref.\,\cite{thermostats}. The
rates were analyzed using a method based on mean first-passage times\,\cite{MFPT}.
The studied $N_{\mathrm{c}}/N$-ratios range from 1 to 20. We have
renormalized the simulation data to the first point of each series
because we are only interested in the deviation arising from an increase
of the amount of carrier-gas. %
\begin{figure}
\includegraphics[width=0.95\columnwidth]{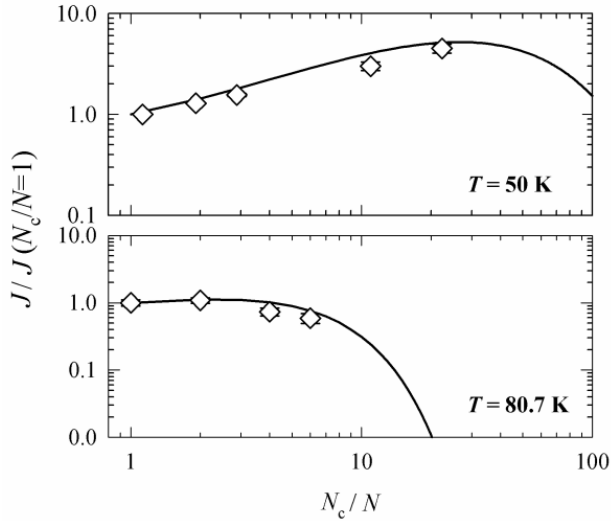}

\caption{\label{fig:Comparison-of-MD}Comparison of MD results with theoretical
prediction of the pressure-effect, Eq.~(\ref{eq:PE_deviation}).}

\end{figure}
Fig.\,(\ref{fig:Comparison-of-MD}) shows the simulation result together
with the prediction of Eq.\,(\ref{eq:PE_deviation}), which is also
renormalized by the same fixed value. The agreement is quite remarkable.
At 50\,K, we only observed a slight increase in the rate with increasing
amount of carrier-gas. On the other hand, the rate already starts
to drop at a much smaller ratio of $N_{\mathrm{c}}/N=4$ at 80.7\,K,
again following the theoretical prediction remarkably close.  

\begin{figure}
\includegraphics[width=0.95\columnwidth]{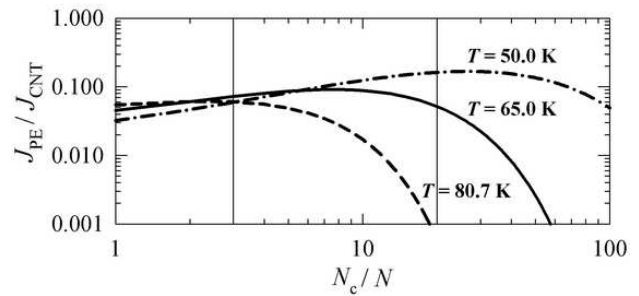}

\caption{\label{fig:var_T}pressure-effect for argon at three different temperatures,
50~K (dash-dot), 65~K (solid), and 80.724~K (dash) as a function
of the ratio of carrier-gas to vapor. The supersaturations at each
temperature correspond to a base rate of about $10^{25}\:\mathrm{cm^{-3}}\mathrm{s}^{-1}$
\cite{Wede_rates}. The vertical lines frame a region of $N_{\mathrm{c}}/N$-ratios
typically encountered in experiments. }

\end{figure}
We now take a closer look at the influence of temperature on the pressure-effect.
Fig.\,(\ref{fig:var_T}) again shows the pressure-effect for argon,
this time for three different nucleation temperatures. We took the
supersaturations at each $T$ to correspond to approx. the same rate
from an earlier work \cite{Wede_rates}. The nonisothermal effects,
Eq.\,(\ref{eq:nonisotherm_rate}), are only weakly depending on the
supersaturation of the system. The $pV$-work term, however, gets
more and more pronounced the higher the temperature. The reason for
this is the huge change in the equilibrium vapor pressure by almost
three orders of magnitude with increasing $T$. Hence, the same $p_{\mathrm{c}}/p$-ratio
corresponds to a much higher total pressure in the system and the
$pV$-work contribution is taking over earlier than at lower $T$.
We have also framed a region of ratios that can be encountered experimentally
\cite{Iland}. In a typical experiment \cite{Brus_et_al,Brus_Accep}
to study the pressure-effect, one would fix $T$ and $S$ and vary
the total pressure by increasing the amount of carrier-gas. We can
do the same in Fig.\,(\ref{fig:var_T}), going from 3:1 to 20:1.
At 50\,K we observe a mild increase of the rate of about a factor
of 2. There is practically no change in the rate at 65\,K, at least
none that would be detectable by available experimental techniques.
Finally, at 80.7\,K, we only observe a strong decrease of the rate
of up to 2 orders of magnitude. Thus, we understand clearly how it
is possible to observe only a positive, a negative, or no effect at
all for the same substance. Both the sign at any given $p_{\mathrm{c}}/p$-ratio
and the magnitude of the effect will strongly depend on the equilibrium
vapor pressure (thus on the substance and temperature) and the experimental
window of accessible $p_{\mathrm{c}}/p$-ratios.

Finally, we discuss some of the generic conclusions that can be drawn
from our model with respect to different experimental situations.
Since experiments greatly vary in the preparation, parameters, rate
window, evaluation method etc., a careful evaluation requires a comprehensive
analysis that we leave for a future work. In general, we can distinguish
between two different limiting behaviors of the pressure-effect, depending
on the relative influence of the $pV$-work term and, therefore, the
total\emph{ }pressure $p_{\mathrm{tot}}=p+p_{\mathrm{c}}$. If $\frac{p_{\mathrm{tot}}v_{l}}{k\mathrm{_{B}}T}\ll\ln S$,
the contribution of $pV$-work can be neglected and the only noticeable
contribution comes from nonisothermal effects. In that case, the possible
influence of the pressure-effect follows the dashed line in Fig.\,\ref{fig:concept}
and the only effect we could observe in an experiment would be a slight
increase in the rate with increasing amount of carrier-gas. Note also
that we would not observe any pressure-effect at all if we already
started from high molar fractions of carrier-gas. Interestingly, these
are the conditions that we typically find in nucleation pulse chambers
\cite{Strey}, where the total pressure varies only very little around
1 bar and where carrier-gas fractions are high. Our analysis thus
justifies why no effect is found in these experiments. In the opposite
limit, when $p_{\mathrm{c}}/p$ is large and $\frac{p_{\mathrm{tot}}v_{l}}{k\mathrm{_{B}}T}\sim\ln S$,
the system is perfectly thermalized but the dominating $pV$-work
leads to a significant overall decrease of the rate. Moreover, we
can make an estimate of the changes in the critical size in terms
of the CNT estimate \cite{Debene},\begin{equation}
n^{*}=\frac{32\pi}{3}\frac{v_{l}^{2}\gamma^{3}}{\Delta\mu_{\mathrm{eff}}^{3}},\label{eq:GT}\end{equation}
again replacing the chemical potential by the effective one, Eq.\,(\ref{eq:eff_chem_pot}).
Eq.\,(\ref{eq:GT}) shows that a further increase of the carrier-gas
pressure (or similarily $p_{\mathrm{tot}}$) would also lead to an
increase in $n^{*}$. In an experiment, this might be noticeable by
an increase in the slope of a nucleation rate isotherm, which is connected
to the critical size via the nucleation theorem \cite{Bowles_2}.
However, care must be taken in that kind of analysis because the nucleation
theorem in principle requires that the carrier-gas pressure $p_{\mathrm{c}}$
is also fixed. 

In summary, we have presented a simple model that is able to shed
light on one of the biggest puzzles in current research of vapor-liquid
nucleation: the {}``pressure-effect''. The investigation of this
effect is often entangled between experimental uncertainties, wildly
different experimental conditions and procedures, and theoretical
interventions on different stages of the modeling of nucleation and
growth. We have presented a physically very appealing way to disentagle
most of these ambiguities by properly incorporating the presence of
a carrier-gas into CNT. Simulation results corroborate the validity
of the model quite impressively. Still, we cannot discard the influence
of other factors on the observed pressure-effect. Nevertheless, these
factors, if applicable, can be added easily to our model as secondary
contributions to the more fundamental and inevitable physical roles
that the presence of an ambient carrier-gas plays in nucleation and
which are accounted for in our model. For example, it is straightforward
to include nonidealities of the vapor and carrier-gas as well as the
compressibility of the liquid, which surely will play a more prominent
role the higher the pressure. Another possible influence we have neglected
here is a change in the kinetic prefactor. Finally, we deliberately
separated the influence of a truly inert carrier-gas pressure from
other effects such as binary nucleation or surface adsorption. It
is somehow misleading to include these under the same {}``pressure-effect''
tag because even though the strength of them may depend on pressure,
their \emph{origin} certainly is not the pressure of the carrier-gas.
In any case, the insights provided by the model presented in this
letter will undoubtedly be very helpful to quantify and remove the
influence of the carrier-gas on experiments. This opens the door to
a more accurate evaluation of nucleation rates, which has important
implications on many atmospheric and technological processes.

\begin{acknowledgments}
This work has been supported by the Atmospheric Composition Change
-- European Network of Excellence (ACCENT), the German Academic Exchange
Service (DAAD), and the Spanish Ministry of Education and Science.
We thank G. Chkonia for assisting on part of the simulations and R.
Strey for valuable discussions.
\end{acknowledgments}


\begin{thebibliography}{10}
\bibitem{Kulmala}M. Kulmala, Science \textbf{302}, 1000 (2003).

\bibitem{turnved}P. Tunved \textit{et al.}, Science \textbf{312},
261 (2006).

\bibitem{Winkler}P. M. Winkler \textit{et al.}, Science \textbf{319},
1374 (2008).

\bibitem{Brus_et_al}D. Brus, V. Ždímal, and F. Stratmann, J. Chem.
Phys. \textbf{124}, 164306 (2006).

\bibitem{Brus_Accep}D. Brus \textit{et al.}, J.Chem. Phys. \textbf{128},
134312 (2008).

\bibitem{Ford}I. J. Ford, J. Aerosol. Sci. 23, \textbf{447} (1992).

\bibitem{Oxt_Laak}D. W. Oxtoby and A. Laaksonen, J. Chem. Phys. \textbf{102},
6846 (1995).

\bibitem{Luij_vDong}C. C. M. Luijten and M. E. H. van Dongen, J.
Chem. Phys. \textbf{111}, 8524 (1999).

\bibitem{Kalikm}V. I. Kalikmanov and D. G. Labetski, Phys. Rev. Lett.
\textbf{98}, 085701 (2007).

\bibitem{Nov_Reiss}V. M. Novikov, O. V. Vasil'ev, and H. Reiss, Phys.
Rev. E \textbf{55}, 5743 (1997).

\bibitem{Debene}P. G. Debenedetti, \emph{Metastable Liquids: Concepts
and Principles} (Princeton University Press, Princeton, 1996).

\bibitem{McGraw_Lett}R. McGraw and A. Laaksonen, Phys. Rev. Lett.
\textbf{76}, 2754 (1996).

\bibitem{Reg_small}D. Reguera \emph{et al}., J. Chem. Phys. \textbf{118},
340 (2003).

\bibitem{Bowles_2}R. K. Bowles \textit{et al.}, J. Chem. Phys. \textbf{113},
4524 (2000).

\bibitem{Feder}J. Feder \emph{et al}., Adv. Phys. \textbf{15}, 111
(1966).

\bibitem{thermostats}J. Wedekind, D. Reguera, and R. Strey, J. Chem.
Phys. \textbf{127}, 064501 (2007).

\bibitem{Iland}K. Iland \emph{et al}., J. Chem. Phys. \textbf{127},
154506 (2007).

\bibitem{merikPRL}J. Merikanto \textit{et al.}, Phys. Rev. Lett.
\textbf{98}, 145702 (2007).

\bibitem{Wede_rates}J. Wedekind \textit{et al.}, J. Chem. Phys. \textbf{127},
154515 (2007).

\bibitem{MFPT}J. Wedekind, R. Strey, and D. Reguera, J. Chem. Phys.
\textbf{126}, 134103 (2007).

\bibitem{Strey}R. Strey, P. E. Wagner, and Y. Viisanen, J. Phys.
Chem. \textbf{98}, 7748 (1994).
\end{thebibliography}
\end{document}